\documentclass[twocolumn,preprintnumbers,amsmath,amssymb]{revtex4-1}
\usepackage{graphicx}
\usepackage{dcolumn}
\usepackage{bm}
\begin{document}
\title{Conductance noise in nano-patterned PbS quantum dot arrays }
\author{Tamar S. Mentzel}
\thanks{These two authors contributed equally to this work.}
\author{Nirat Ray}
\thanks{These two authors contributed equally to this work.}
\affiliation{Department of Physics, Massachusetts Institute of Technology, Cambridge, Massachusetts 02139, U.S.A.}
\author{Darcy D. W. Grinolds}
\affiliation{Department of Chemistry, Massachusetts Institute of Technology, Cambridge, Massachusetts 02139, U.S.A.}
\author{Neal E. Staley}
\affiliation{Department of Physics, Massachusetts Institute of Technology, Cambridge, Massachusetts 02139, U.S.A.}
\author {Moungi G. Bawendi}
\affiliation{Department of Chemistry, Massachusetts Institute of Technology, Cambridge, Massachusetts 02139, U.S.A.}
\author{Marc A. Kastner}
\email{mkastner@mit.edu}
\affiliation{Department of Physics, Massachusetts Institute of Technology, Cambridge, Massachusetts 02139, U.S.A.}

\date{\today}
\begin{abstract}
We report unexpectedly large noise in the current from nanopatterned PbS quantum dot films. The noise is proportional to the current when the latter is varied by changing the source-drain bias, gate voltage or temperature. The spectral density of the noise is given by a power law in frequency at room temperature, but remarkably, we observe a transition to telegraph noise at lower temperatures. The probability distribution of the off-times follows a power law, reminiscent of fluorescence blinking in colloidal quantum dot systems. Our results are understood simply in terms of conductance fluctuations in a quasi-one dimensional percolation path, and more rigorously in terms of a model in which charge through the film is transmitted in discrete time intervals, with the distribution of  intervals completely described by Levy statistics. 
\end{abstract}
\maketitle

Quantum dots made in various ways have received great attention because their properties can be controlled by their size. They are artificial atoms, for which the charge and excitation energies are quantized and determined by their dimensions. Transport measurements on individual dots have revealed Coulomb blockade \cite{kastner1992single,klein1997single} and fascinating many-body phenomena including the Kondo effect \cite{kastner2001kondo}. Colloidal quantum dots can be easily made to self assemble into two and three dimensional arrays, and their unique optical properties have already found application in displays and light-emitting diodes \cite{shirasaki2013emergence}. They are attractive for application in photovoltaics and optoelectronics because they are low-cost and solution processable \cite{talapin2009prospects}.  Additionally, their predicted unconventional charge and spin transport properties hold promise for quantum computation and spintronics \cite{ouyang2003coherent}.  However, their electronic transport properties are still poorly controlled, compared to their optical properties.  If one were able to control the tunneling rates between quantum dots in an array, one would have a system described by a tunable Hubbard model, just as the single-electron transistor is described by a tunable Anderson model that underlies the Kondo effect, thus providing a platform for a deeper insight into correlated electron physics.

Until now, studies of charge transport in quantum dot arrays have revealed signatures of disorder, giving limited insight into the electronic properties intrinsic to the quantum dots \cite{PhysRevB.77.075316,morgan2002electronic,ginger2000charge}. A careful study of noise in a material can provide additional insights into its properties, and can even be used to characterize exotic states such as charge density waves \cite{cox2008sliding} and quasi-particles in a fractional quantum hall state \cite{bid2010observation}. Noise studies in colloidal quantum dot systems have been primarily focused in two directions. Because of the effort to integrate colloidal dots into devices, the first direction has been to understand 1/f noise in the granular arrays that may impact device performance \cite{lai2014low,liu20141,lhuillier2012colloidal,keuleyan2011mid}. The second direction is the study of fluorescence intermittency or blinking in colloidal dots, where the dots alternate between on (fluorescing) and off (non-fluorescing) states. More recently, blinking behavior has been observed from individual core/shell PbS/CdS and InAs/CdSe dots \cite{correa2012single},
suggesting that the processes governing stochastic blinking in colloidal dots are universal, and insensitive to inherent material properties. However, to our knowledge, transport signatures of blinking have never been observed.

\begin{figure}
\includegraphics[scale =1.0]{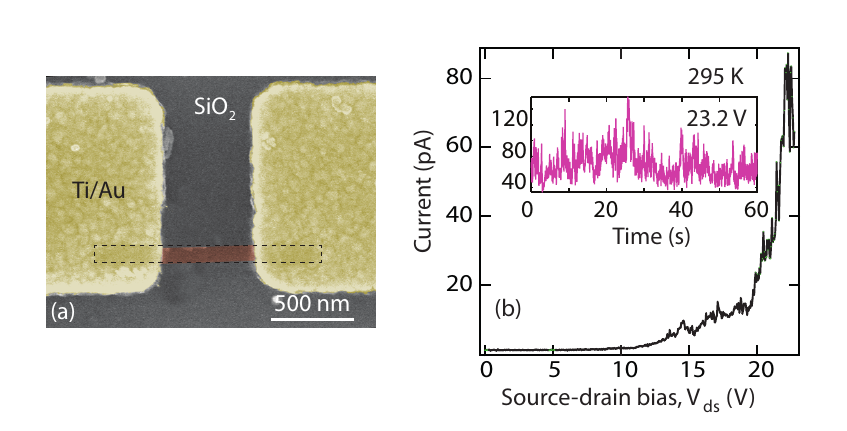}
\caption{(Color online) (a) False color scanning electron micrograph of a device similar to those used for current measurements on butylamine-capped PbS dot films. The nanopattern is 80 nm wide and 600 nm long (the black rectangle overlapping the Ti/Au is the nanopattern).  (b)  Current as a function of the voltage between the Ti/Au electrodes (drain and source) measured at room temperature.  Inset shows the current as a function of time for V$_{ds}$ = 23.2 V. Note that the fluctuations of the current occur in the time domain and are as large as approximately 30\% of the current.}
\label{fig1}
\end{figure}

In this letter, we present measurements of current noise in nanopatterned PbS colloidal quantum dot assemblies. We use nanolithography to make patterns estimated to be as small as 12 dots wide. We find very large and unexpected noise in the current through them. The noise is proportional to the current, consistent with what is expected from conductance rather than charge-density fluctuations, and it increases as the width of the sample decreases. In some cases, particularly at lower temperatures, the noise is telegraph-like, switching suddenly between two states. The off times become larger with time leading to an overall decay of the current, and we find that the current can be restored by heating the sample to an elevated temperature. All of these observations are consistent with a model, in which charge transport through the film occurs in discrete intervals through a quasi one dimensional percolation path.

We use PbS dots with a mean diameter of approximately 4.5 nm. The dots are synthesized air-free using a Schlenk line by high temperature pyrolysis of Pb and S precursors in an oleic-acid/octadecene mixture \cite{steckel20031,zhao2010colloidal}. We process the growth solution to remove remaining products and, if desired, perform a solution exchange of the native oleic acid capping ligand for a smaller molecule, n-butylamine. For electric current measurements we create a field-effect structure, using doped Si as the gate and 100-300 nm of SiO$_2$ as the gate insulator. The source and drain electrodes are made of 7 nm/50 nm Ti/Au, and patterned using electron beam lithography. We create the nanopatterned PbS films using a novel technique based on electron beam lithography followed by a lift-off process as detailed in \cite{mentzel2012nanopatterned}. Nanopatterning allows us to minimize macroscopic structural defects such as cracking and clustering, and enables the measurement of properties intrinsic to the colloidal dots. For some of our measurements the Ti/Au electrodes are much wider than the dot array, as shown in Fig. \ref{fig1}(a), but in other cases they are as narrow as the 80-200 nm wide nanopattern.  Sample processing occurs in the inert glovebox environment, and the dots are only briefly exposed to air while wirebonding and loading the cryostat. The sample is thereafter always kept in high purity He or vacuum. We perform two-point DC measurements to study the noise and current voltage characteristics of the film, using a Femto low noise current amplifier and a NI-6110 high speed voltage card.

\begin{figure}
\includegraphics[ scale =1.0]{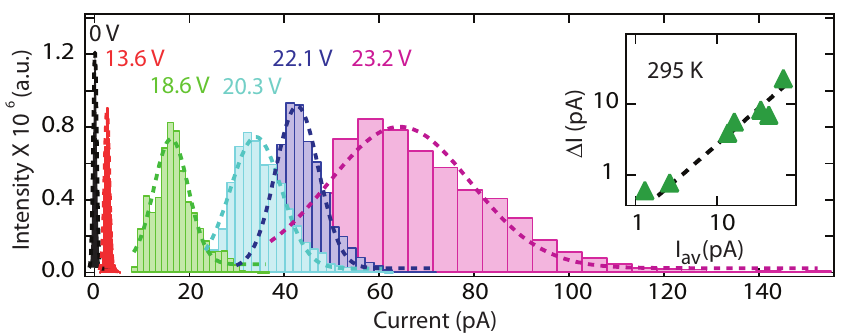}
\caption{(Color Online) Histogram of the current measured as a function of time at 295 K for different values of the source drain bias, as indicated in the figure. The dashed lines in the histograms are Gaussian fits. The current noise, $\Delta$I, is given by the full width at half maximum from the fits. Inset shows log-log plot of the current noise, $\Delta$I, as function of the average current, I$_{av}$, above the noise floor of our circuit. The current noise, $\Delta$I, is linearly  proportional to the average current, I$_{av}$, as seen from the slope near unity on the log-log plot. The black dashed line is a fit to $\Delta$I = 0.27 I$_{av}$. }
\label{fig2}
\end{figure}

A plot of current versus voltage is shown in Fig. \ref{fig1}(b) for a film of n-butylamine-capped dots that is 80 nm wide and 600 nm long.  Because the dot diameter is 4.5 $\pm$ 0.2 nm and the spacing between dots because of the cap layer is $\sim$ 0.6 nm, assuming a cubic packing of the dots, we estimate that the film consists of layers that are about 15 dots wide and 120 dots long. From atomic force microscope (AFM) measurements of the nanopattern height, we estimate that the film is approximately 4 monolayers thick.  The current is below the noise limit of our amplifier for voltages less than 10 V, but at higher voltages we observe large fluctuations in the current. 

To determine how the noise varies with current when the voltage between the source and drain, V$_{ds}$, is varied, we fix the voltage and monitor the current as a function of time. The inset in Fig. \ref{fig1}(b) shows the observed fluctuations as a function of time. For these time traces, we sample the current every 0.01 s, corresponding to an effective band width of 100 Hz.  We histogram the current measured for relatively short timescales (60-600 s) at different values of source drain bias, as shown in Fig. \ref{fig2}. For these short timescales, each histogram can be approximately described by a Gaussian distribution, and we estimate the average fluctuation size, $\Delta$I, from the full-width at half maximum.  The inset of Fig. \ref{fig2} shows a plot of $\Delta$I as a function of the average current, I$_{av}$, above the limit given by our circuit noise. From the the slope near unity on the log-log plot, we find that the current noise is linearly proportional to the average current, consistent with what is expected for conductance rather than charge-density fluctuations.

\begin{figure}
\includegraphics[ scale =1.0]{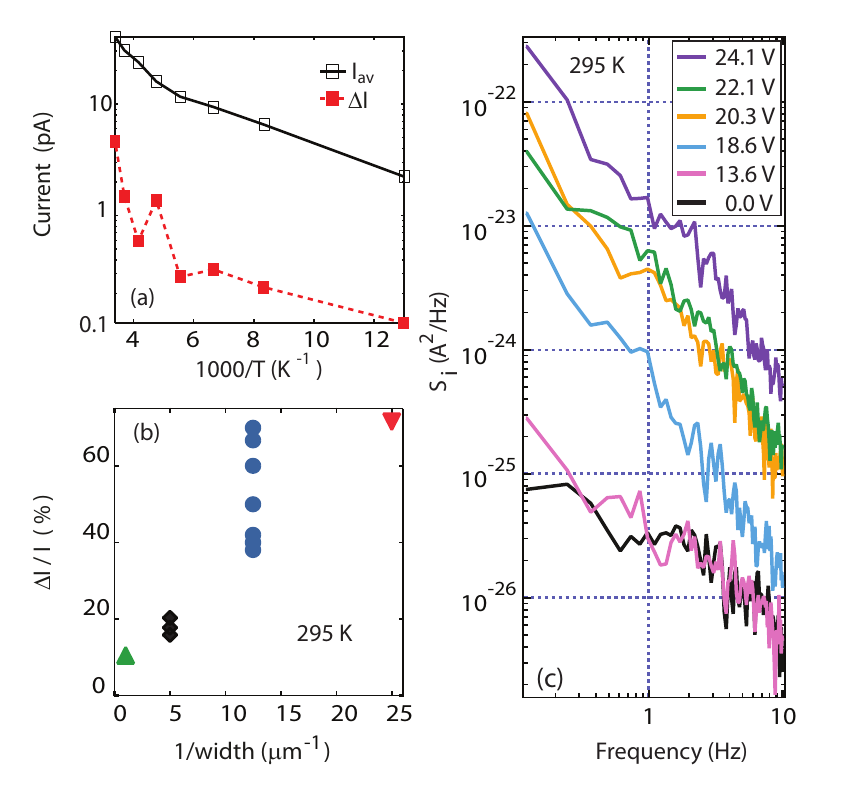}
\caption{(Color Online) (a) Measurement (with a different but identically prepared sample) of $\Delta$I and I as functions of inverse temperature at V$_{ds}$ = 15 V. That the curves are roughly parallel indicates that $\Delta$I/I is constant as T is varied.(b) Relative current noise $\Delta$I/I for nanopatterns of different width.  At a width of 80 nm, we show the sample-to-sample variation.  (c) The power spectral density of the noise, for different voltages, showing that it has an approximate 1/f$^\gamma$ dependence on frequency, f, below 10 Hz, with $\gamma$ equal to 1.4 $\pm$ 0.1. }
\label{fig3}
\end{figure}

We have also studied the noise as a function of temperature and gate voltage, and find that the proportionality holds irrespective of how the current is changed. This is illustrated in Fig. \ref{fig3}(a) for the temperature variation, where we show that $\Delta$I decreases with temperature in a way that is approximately proportional to I. 

This large excess noise has not been observed previously on large scale patterns because $\Delta$I/I grows with decreasing pattern size. This is illustrated in Fig. \ref{fig3}(b) where we plot $\Delta$I/I as a function of inverse sample width. The wider patterns are made identically and have a thickness variation of $\pm$ 20 nm from sample to sample. Despite the large sample-to-sample variation, illustrated for samples of width approximately 80 nm, the general trend, that the noise is smaller for wider samples, is clear.

At room temperature, we find the power spectral density of the noise at low frequency to be of the form 1/$f^\gamma$, with $\gamma = 1.4$ $\pm$ 0.1. Figure \ref{fig3}(c) shows this dependence and also how the amplitude of the 1/$f^\gamma$ noise grows with increasing source-drain bias, while the exponent $\gamma$ remains almost unchanged. Our observation of large noise even when cracks and clusters are minimized by nanopatterning, suggests an intrinsic origin of the noise.

\begin{figure}
\includegraphics[ scale =1.0]{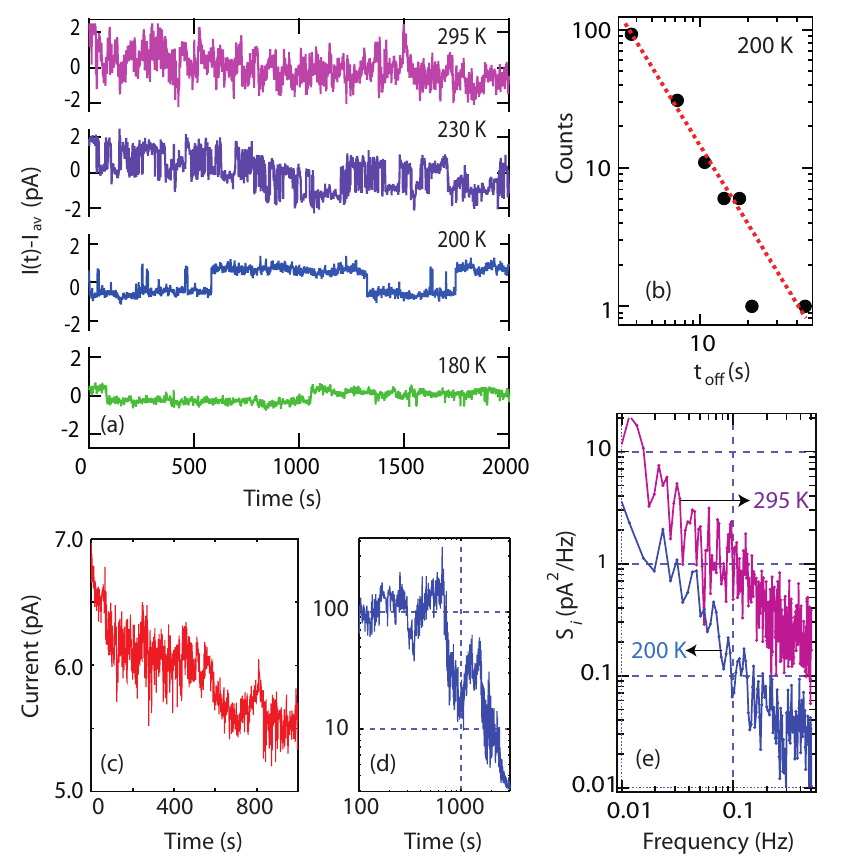}
\caption{(Color Online) (a) Current as a function of time in nanopatterned arrays of oleic-acid-capped PbS dots showing telegraph noise. (b) Histogram of the off times at 200 K fit to a power law, P(t$_{off}$)$\propto$1/t$_{off}^{1+\mu}$, with $\mu$ = 0.6 $\pm$ 0.15. (c) Observed decay of the current immediately after setting the voltage to 8 V at room temperature. The decay can be approximated by a power law, I = I$_o$t$^{-\alpha}$, with $\alpha$ = 0.4 $\pm$ 0.1. (d) Log-log plot of the largest observed decay of the current, measured in n-butylamine-capped PbS dots. Power law fit to the data gives $\alpha$ = 1.0 $\pm$ 0.2. (e) Spectral density of noise at room temperature and at 200 K where discrete telegraph events are seen showing a 1/f$^\gamma$ spectrum, where $\gamma$ is comparable to $1+\mu$ for the wait time distributions.}
\label{fig4}
\end{figure}

The general features of the noise remain unchanged for PbS dots with a different capping molecule or ligand. We perform noise measurements on 200-nm-wide nanopatterns of oleic-acid-capped PbS dots. Though the patterns are wider, if we take into account the length of the native ligands, the number of dots are comparable to the 80-nm-wide nanopatterns made with n-butylamine-capped dots. 
Figure \ref{fig4}(a) shows traces of the current as a function of time with a fixed applied voltage of 9 V at various temperatures.  We see clear telegraph noise with switching rates that increase with increasing temperature. The relative amplitude of the telegraph switches also increases as the current is increased by going to higher temperatures. We have observed telegraph switches in n-butylamine-capped dots as well, particularly at lower temperatures ($<$ 100 K). They are, however, seen much more clearly in the current from oleic-acid-capped dots, and at higher temperatures.  In most cases where switching noise has been observed in the past, such as in the conductance of mesoscopic MOSFETS \cite{ralls1984discrete}, a histogram of the off (lower current) times or the on (high current) times shows an exponential decay.  However, as illustrated in Fig. \ref{fig4}(b), a histogram of the off times at 200 K is a power law with an exponent of 1.6 $\pm$ 0.15.  

The existence of a power law distribution of wait times also means that the more time passes, the more likely it is that the current is in the low state. As the rate of the current decay is scaled by the magnitude of the average current, it is observed more prominently at higher temperatures. This is shown in Fig. \ref{fig4}(c) where the average current clearly decreases with time. The observed decay can be fit to a power law, t$^{-\alpha}$, where $\alpha$ = 0.4 $\pm$ 0.1.  Figure \ref{fig4}(d) shows the largest observed current decay with an exponent $\alpha$ = 1.0 $\pm$ 0.2, measured in n-butylamine-capped dots. The magnitude of the decay exponent varied from sample to sample, but has been observed to satisfy 0  $< \alpha <$ 1, within the error bar of the measurements.

The spectral density of the noise at higher temperatures is again of the form 1/f$^\gamma$, with $\gamma$ = 1.4 $\pm$ 0.1, as shown in Fig. \ref{fig4}(e). At lower temperatures where we see discrete switching events with a power law distribution of wait times, the noise spectrum is also a power law in frequency with a similar exponent, as opposed to a white spectrum expected for poissonian processes \cite{novikov2005levy}. 

The very large size of $\Delta$I/I, its increase with decreasing sample width and the observation of telegraph noise strongly suggests that the current in the film is carried by quasi-one-dimensional channels.  The quantity $\Delta$I/I approaches 70\% at a width of 40 nm, suggesting that this is approximately the typical separation of these channels for the n-butylamine-capped dots, or about 10 times the diameter of the dots.  Following the arguments of Ambegaokar, Halperin and Langer (AHL) \cite{ambegaokar1971hopping}, for variable range hopping in disordered semiconductors, we should not be surprised at this.  The high resistance of the barriers between dots results in exponentially small tunneling rates between them, and small variations of tunnel barrier heights and widths will therefore result in exponentially large variations in tunneling resistances.  The AHL recipe for calculating the overall resistance with such an exponentially broad distribution of microscopic resistances is to begin with the smallest resistors and put in successively larger ones until a percolation path is created across the sample.  Such a path is necessarily quasi-one-dimensional.  If a resistor in the critical path switches between two values of resistance and if there are only a small number of parallel paths in the sample, telegraph noise will be seen.  

The power-law dependence of the off times in the telegraph noise is also known as Levy statistics.  This behavior has been observed in the blinking of the photoluminescence of individual quantum dots  \cite{shimizu2001blinking,chen2013compact}, and is typically understood by the existence of trap states that facilitate non-radiative recombination \cite{zhao2010challenge, efros2008nanocrystals}.  We assume that trapping of charge also strongly modifies the resistance along the percolation path.  This could happen, for example, if the trapped charge alters a tunnel barrier along the path, and a power law dependence of the off time could result from a distribution of trap depths. 

The hypothesis that trapping is the cause of the overall decay of the current is substantiated by raising the temperature after the average current is allowed to decay.  We find that such annealing restores much of the lost current. To quantify this effect, we measure the current-voltage characteristics of a 200-nm-wide n-butylamine-capped PbS nanopattern at room temperature and subsequently heat the sample under vacuum to 303 K, while continuously monitoring the current. From the temperature dependence of the current at lower temperatures, we expect the current to increase by a few percent as the temperature is raised to 303 K; instead we observe that the current almost doubles with heating. Further controlled experiments are needed to quantify the observed increase in current with annealing.

Novikov \textit{et al.} \cite{novikov2005levy} developed a model for the time decay of the average current first observed in CdSe dots capped with trioctylphosphine oxide \cite{morgan2002electronic}. The authors considered the array to be made of a finite number of conducting channels, and showed that both a decay of the average current in a single channel and 1/f noise can result from a power-law tail for the distribution of off-times ($\tau$), given by:
\begin{equation}
p(\tau) = \frac{A}{\tau^{1+\mu}},  0<\mu<1.
\end{equation}

The decay of the average current  is characterized by the exponent $\alpha$, where $\alpha$ = 1 - $\mu$ and 0 $<$ $\alpha$ $<$ 1. Subsequently, it has been suspected that the behavior is not generalizable, because the CdSe films are found \cite{morgan2002electronic} to have blocking contacts to Ti/Au, the decay of the current might therefore be a contact effect. From our measurements, if we assume a power law decay of the current, the decay exponent $\alpha$ = 0.4 $\pm$ 0.1, is consistent with $\mu$ $\sim$ 0.6 from the distribution of wait times.  

The results reported here suggest that the limitation of the current by quasi-one-dimensional channels and its noise and decay may be general features of colloidal quantum dots.  Our results also suggest that charge is transmitted in discrete time intervals, described by Levy statistics.  This insight into the mechanism of charge transmission in a film of colloidal dots is an essential step toward accessing the predicted complex electron and spin behavior in quantum dot films. Furthermore, understanding the electrical properties of the films unimpeded by structural defects will make it possible to optimize the efficiency of nanocrystal-based devices and applications.


We acknowledge useful discussions with Prof. Leonid S. Levitov. We are grateful to Mark Mondol and the RLE SEBL facility for
experimental help with electron-beam lithography. This work was supported in part by the U. S. Army Research Laboratory and the U. S. Army Research Office through the Institute for Soldier Nanotechnologies, under contract number W911NF-13-D-0001 (synthesis and nano-patterning of colloidal quantum dots), by the Department of Energy under award number DE-FG02-08ER46515 (transport and noise measurement on colloidal quantum dots, including device fabrication) and by Samsung SAIT. NR acknowledges support from the Schlumberger Foundation through the Faculty for the Future Fellowship Program. DDWG acknowledges support from the Fannie and John Hertz Foundation Fellowship.  

\bibliography{Bib}





\end{document}